\documentclass[a4paper,11pt]{article}
\usepackage{jcappub}
\usepackage{lineno,mathrsfs}
\modulolinenumbers[5]
\usepackage{amsmath,amssymb,epsfig,bm,mathrsfs}
\usepackage{color,slashed,exscale,multirow,soul}
\usepackage{times,txfonts,xcolor,graphicx,tikz}
\usepackage{xspace}  
\usepackage[normalem]{ulem}

\newcommand{\MR}{$M$-$R$\xspace}

\usepackage[normalem]{ulem}  

\newcommand{\bea}{\begin{eqnarray}}
\newcommand{\eea}{\end{eqnarray}}

\newcommand{\MeV}{\text{MeV}}
\newcommand{\fm}{\text{fm}}
\newcommand{\tr}{\text{tr}}
\newcommand{\cs}{\ensuremath{c_\text{s}}}
\newcommand{\csq}{\ensuremath{c_\text{s}^2}}
\newcommand{\Lsym}{\ensuremath{L_{\text{sym}}}}

\newcommand{\Qsat}{\ensuremath{Q_{\text{sat}}}}

\arxivnumber{2409.05322} 
\title{\LARGE Confronting new NICER mass-radius measurements with 
phase transition in dense matter and \\ twin compact stars}
\author[a]{Jia Jie Li}
\author[b,c]{Armen Sedrakian}
\author[d]{Mark Alford}
\affiliation[a]{School of Physical Science and Technology, 
Southwest University, Chongqing 400715, China}
\affiliation[b]{Frankfurt Institute for Advanced Studies,
D-60438 Frankfurt am Main, Germany}
\affiliation[c]{Institute of Theoretical Physics,
University of Wroc\l{}aw, 50-204 Wroc\l{}aw, Poland}
\affiliation[d]{Department of Physics, Washington University,
St.~Louis, Missouri 63130, USA}
\emailAdd{jiajieli@swu.edu.cn; sedrakian@fias.uni-frankfurt.de; alford@physics.wustl.edu}
\abstract{
\newline
The (re)analysis of data on the X-ray emitting pulsars PSR J0030+0451 
and J0740+6620, as well as new results on PSR J0437-4715 and J1231-1411, 
are confronted with the predictions of the equation of state (EoS) 
models allowing for strong first-order phase transition for the mass-radius 
($M$-$R$) diagram. We use models that are based on a covariant density 
functional (CDF) EoS for nucleonic matter at low densities and a quark 
matter EoS, parameterized by the speed of sound, at higher densities. 
To account for the variations in the ellipses for PSR J0030+0451 obtained 
from different analyses, we examined three scenarios to assess their 
consistency with our models, focusing particularly on the potential 
formation of twin stars. We found that in two scenarios, where the 
ellipses for PSR J0030+0451 and J0437-4715  with masses close to 
the canonical mass $\sim 1.4\,M_{\odot}$ are significantly separated, 
our models allow for the presence of twin stars as a natural explanation 
for potential differences in the radii of these stars.
}

\begin{document}
\maketitle

\flushbottom

\section{Introduction}
The NICER observations of nearby neutron stars allowed for accurate
(up to 10\%) inferences of neutron star radii in conjunction with the
masses of nearby X-ray-emitting neutron stars. Recent (re)analysis of
the data of four millisecond pulsars - the two-solar-mass pulsar PSR
J0740+6620 (hereafter J0740) and two canonical mass $1.4\,M_{\odot}$
stars PSR J0437-4715 (hereafter J0437) and J0030+0451 (J0030), 
and the one-solar-mass pulsar PSR J1231-1411 (J1231) pose a
challenge to the modern theories of dense matter to account for the
features observed on the mass-radius ($M$-$R$) diagram of neutron
stars. A sharp first-order transition between hadronic and quark
matter can produce a disconnected branch of hybrid stars, opening up
the possibility of {\em twin} stars, where there are two
different stable configurations, with different radii, but having
the same mass. The larger star will be composed entirely of
hadronic matter, while the more compact star will be a hybrid star
with a quark core in the central
region~\cite{Glendenning:2000,Zdunik:2013,Benic:2015,Alford:2017,
AlvarezCastillo:2019,Blaschke:2020,Christian:2022,Lijj:2020a,Lijj:2021,
Lijj:2023a,Lijj:2023b,Laskos-Patkos:2023,Naseri:2024,
Jimenez:2024,Zhang:2024,Kini:2024,Laskos-Patkos:2024},
for recent reviews see
Refs.~\cite{Baym:2018,Sedrakian:2023part}. Between the hadronic branch 
and the hybrid branch there is a range of radii for which there are no 
stable configurations. This is true if the transition from hadronic to 
quark phase is rapid compared to other time scales in the problem, 
for example, the period of fundamental modes by which these stars become 
unstable. In the case of slow conversion, the stability is 
recovered~\cite{Rau:2023PRD_a,Rau:2023PRD_b,Lenzi:2023PRD,Ranea-Sandoval:2023PRD,Rather:2024JCAP}.

Recently, new and updated NICER astrophysical constraints have been
published for the four pulsars
mentioned~\cite{Vinciguerra:2024,Salmi:2024a,Salmi:2024b,Choudhury:2024}. 
Notably, the analysis of PSR J0030 resulted in three different ellipses in
the \MR plane, each corresponding to a different analysis model. 
For PSR J1231, the inference results obtained with the preferred PDT-U model 
show a strong sensitivity to the choice of radius priors, with stable and 
likely converged outcomes achieved only under constrained radius priors~\cite{Salmi:2024b}.
The purpose of this paper is to assess the compatibility of
these new analyses with the hybrid star models we recently developed
in a series of
papers~\cite{Lijj:2024,Lijj:2023a,Lijj:2023b}. Specifically, we will
examine three scenarios, labeled A, B, and C, which share the same
data for PSR J0437 and PSR J0740 but incorporate different analyses for 
PSR J0030 using different models of the surface temperature patterns. 
The scenarios are defined in Table~\ref{table:data}.
In model ST-U each of the two hot spots is described by a single spherical cap; 
in CST+PDT there is a single temperature spherical spot with two components, 
one emitting and one masking;
in ST+PST the primary (ST) is described by a single spherical cap and the 
secondary (PST) by two components, one emitting and one masking; 
in ST+PDT the primary (ST) is described by a single spherical cap and the 
secondary (PDT) by two components, both emitting; 
in PDT-U each of the two hot spots is described by two emitting spherical caps.
For details see Refs.~\cite{Vinciguerra:2024,Choudhury:2024}.

In confronting the models of hybrid stars we will pay special
attention to the possibility of twin stars in Scenario C, as in this 
case the data for canonical mass pulsars J0439 and J0030 does not overlap
at $2\sigma$ ($95\%$ confidence) level, hinting towards the existence of 
twin configurations. For this to occur, a strong first-order phase 
transition is needed, which will be parametrized in terms of the fractional 
energy density jump $\Delta \epsilon/\epsilon_\text{tr}$.

\begin{table}[tb]
\centering
\setlength{\tabcolsep}{5.0pt}
\begin{tabular}{ccccccc}
\hline\hline
\multirow{2}*{Scenario}& J0740 &  J0437 & J1231  &       &  J0030 &        \\
\cline{5-7}
                       &  ST-U & CST+PDT& PDT-U  & ST+PST& ST+PDT & PDT-U  \\
\hline
A        & $\times$ & $\times$ & $\times$ & $\times$ &          &          \\
B        & $\times$ & $\times$ & $\times$ &          & $\times$ &          \\
C        & $\times$ & $\times$ & $\times$ &          &          & $\times$ \\
\hline\hline
\end{tabular}
\caption{
New NICER astrophysical constraints~\cite{Vinciguerra:2024,Salmi:2024a,Salmi:2024b,Choudhury:2024} 
used for the three scenarios A, B, C in the present work.
}
\label{table:data}
\end{table}

The remainder of this work is organized as follows. We briefly describe 
the physical foundations of the equation of state (EoS) models used in this 
study in section~\ref{sec:Models}. The theoretical stellar models are 
compared with astrophysical observations in section~\ref{sec:Astro}. 
Finally, our conclusions are presented in section~\ref{sec:Conclusions}.

\section{Models of hybrid stars}
\label{sec:Models}
To ensure this presentation is self-contained, we briefly review the
setup from Li et al.~\cite{Lijj:2024}. We use four representative
nucleonic EoS models based on CDF
theory, as discussed in Oertel et~al.~\cite{Oertel:2017} and Sedrakian
et~al.~\cite{Sedrakian:2023}. Our models are part of the DDME2 family
introduced by Lalazissis et~al.~\cite{Lalazissis:2005}, but feature
varying slope $\Lsym$ of symmetry energy. Our models are labeled as
DDLS-$\Lsym$, see Ref.~\cite{Lijj:2023c}. We keep the skewness
constant at the value implied by the parameterization of
Ref.~\cite{Lalazissis:2005}, $\Qsat=479.22\,\MeV$. Our models share
the following nuclear matter parameters: Saturation density
$\rho_\text{sat}=0.152\,\fm^{-3}$; Binding energy per particle
$E_\text{sat}=-16.14\,\MeV$ at saturation density; Incompressibility
$K_\text{sat}=251.15\,\MeV$; Symmetry energy
$E_\text{sym} =27.09\,\MeV$ at the crossing density
$\rho_\text{c}=0.11\,\fm^{-3}$.

In our analysis below we select from the family of DDLS-$\Lsym$ two
stiff EoS with values of $\Lsym= 80,\,100\,\MeV$ and two soft EoS with
$\Lsym= 40,\,60\,\MeV$ as representative EoS on the basis of which we
built our hybrid star models.  Note that the EoS of neutron-rich
matter below and around $2\,\rho_\text{sat}$ is dominated by the
isovector parameters of a CDF, see for instance
Ref.~\cite{Lijj:2023e}, which in the present setup is encoded in
$L_\text{sym}$. The variations of isoscalar parameters, such as
$K_\text{sat}$ or $Q_\text{sat}$, affect the high-density part of the
EoS and are unimportant because in our models the hadron-quark
phase transition takes place at lower densities, in the range of
$2 \leqslant \rho / \rho_\text{sat} \leqslant 2.5$.

The first-order transition to quark matter is parameterized by
the baryon density and energy density ($\rho_\tr$ and
$\epsilon_\tr$) at which the transition occurs, the energy density
jump $\Delta\epsilon$, and the speed of sound $\cs$ in the quark
matter phase. Following Ref.~\cite{Lijj:2024}, we first fix
$\rho_{\tr}$ and $\epsilon_{\tr}$, and then vary $\Delta \epsilon$
and $\cs$. As described in Ref.~\cite{Alford:2013}, we can identify
specific values of these parameters that yield a \MR relation with
two disconnected branches. The stars with low central density are
purely hadronic, while those with high central density contain a
quark core (i.e., hybrid stars) and are separated by an unstable
region. This topology could lead to twin configurations, where
stars with identical masses differ in their geometric properties
such as radii, moments of inertia, and tidal deformabilities.

\begin{figure*}[tb]
\centering
\includegraphics[width = 1.0\textwidth]{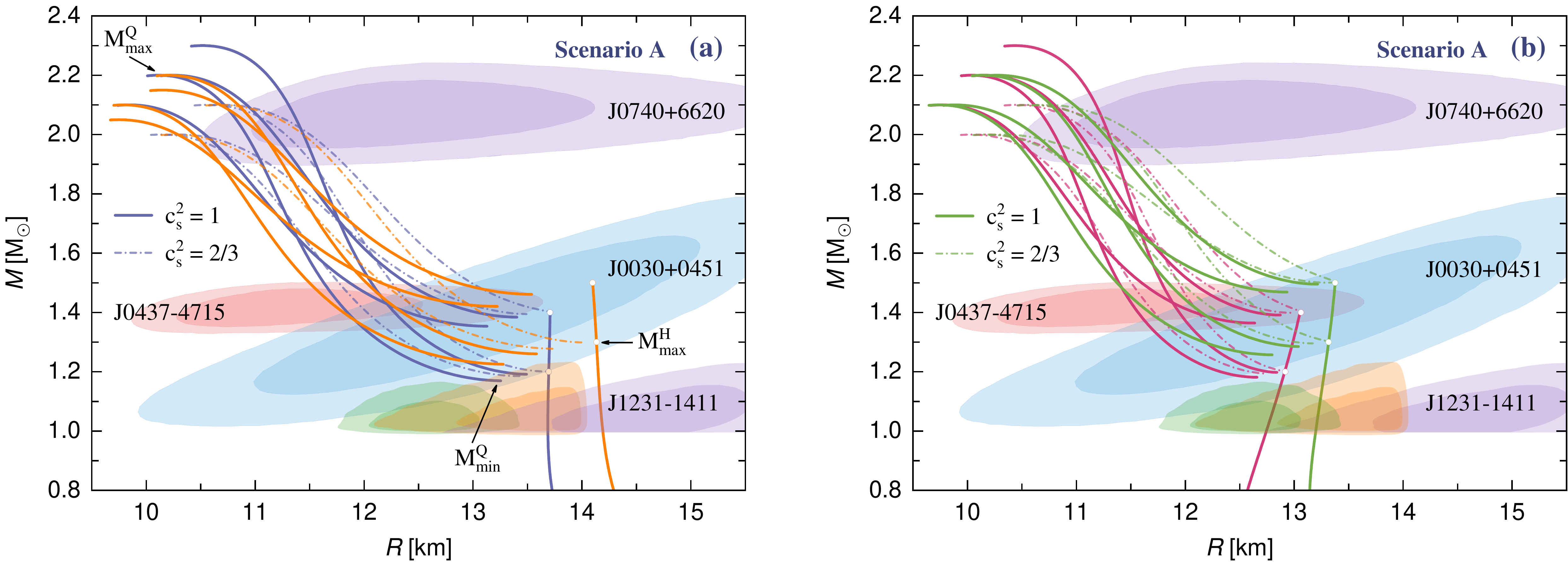}
\caption{
\MR relations for hybrid EoS models with transition mass 
$M^{\rm H}_\text{max}$ in the range $1.2$-$1.5\,M_{\odot}$ and maximum mass 
$M^{\rm Q}_\text{max}$ in the range $2.0$-$2.3\,M_{\odot}$ that were 
constructed from four representative nucleonic EoS models. Models with 
stiff nucleonic EoS (DDLS-80 and DDLS-100) are shown in panel (a), those 
with soft nucleonic EoS (DDLS-40 and DDLS-60) in panel (b).  
Ellipses show observation constraints at 68\% and 95\% credible levels 
from analysis of NICER observations according to 
Refs.~\cite{Vinciguerra:2024,Salmi:2024a,Salmi:2024b,Choudhury:2024}.
The light blue ellipse corresponds to ST+PST analysis 
for PSR J0030~\cite{Vinciguerra:2024}, 
the three regions for J1231 correspond to PDT-U analysis with 
three different radius priors~\cite{Salmi:2024b}.
}
\label{fig:MR_a}
\end{figure*}

For convenience, we use astrophysical parameters instead of microscopic 
ones like $\rho_\text{tr}$ and $\Delta\epsilon$ to characterize a hybrid 
EoS model, see for illustration panel (a) of Figure~\ref{fig:MR_a}. 
These macroscopic parameters are:
(a) $M^\mathrm{H}_\text{max}$: the maximum mass of the hadronic branch;
(b) $M^\mathrm{Q}_\text{max}:$ the maximum mass of the hybrid branch; 
(c) $M^\mathrm{Q}_\text{min}$: the minimum mass of the hybrid branch. 
The last quantity allows us to determine the range of masses where twin 
configurations exist.

As in Ref.~\cite{Lijj:2024} we use the constant speed of sound (CSS) 
parameterization for the quark EoS, which assumes that the 
speed of sound in quark matter remains constant over the relevant density 
range. In the extreme high-density limit we expect $\csq \to 1/3$ which 
corresponds to the conformal limit describing weakly interacting massless 
quarks. The maximum possible speed of sound is the causal limit
$\csq = 1$. Intermediate values represent varying degrees of stiffness 
in the quark matter EoS. Below we will explore two possibilities, 
$\csq = 1$ and 2/3. 

The maximally stiff EoS yields the largest maximum masses for
hybrid stars and the maximal difference in the radii of twin
stars. The intermediate value is more realistic and mimics what one
might expect for non-perturbatively interacting quark matter.
Although we will restrict our analysis to the basic CSS model
with a single speed of sound over the relevant density range, one
could perform a more general analysis using different speeds of
sound in different density segments; it is known that this can
produce triplets of stars, i.e., three stars with equal masses but
different radii~\cite{Alford:2017,Lijj:2020a}.  

\section{Comparison with astrophysical observations}
\label{sec:Astro}
We examine hybrid EoS models with transition masses 
($M^\mathrm{H}_\text{max}$) ranging from 1.2 to 1.5 solar masses. 
This range encompasses the mass estimation of PSR J0437 reported by 
Choudhury et~al.~\cite{Choudhury:2024}. Our analysis aims to evaluate 
the hypothesis that PSR J0437 and J0030 could be twin stars, the first 
being hadronic and the second hybrid.

We use the following mass and radius measurements:
\begin{itemize}
\item PSR J0437-4715: We use the first mass and radius estimates for
  this brightest pulsar by using the 2017-2021 NICER X-ray
  spectral-timing data from Choudhury et~al.~\cite{Choudhury:2024}.
  The preferred CST+PDT model used informative priors on mass,
  distance and inclination from PPTA radio pulsar timing data and
  took into account constraints on the non-source background and validated 
  against XMM-Newton
  data~\cite{Choudhury:2024}.
\item PSR J0030+0451: We use three alternative mass and radius
  estimates from the reanalysis of 2017-2018 data, as reported by
  Vinciguerra et~al.~\cite{Vinciguerra:2024}. One of the estimates 
  is based on the ST+PST NICER-only analysis of the data reported 
  in~\cite{Riley:2019}, but uses an improved analysis pipeline and 
  settings. The two other estimates are based on the joint analysis 
  of NICER and XMM-Newton data which are labelled as ST+PDT and PDT-U. 
  The ST+PDT results are more consistent with the magnetic field geometry 
  inferred for the gamma-ray emission for this 
  source~\cite{Kalapotharakos:2021,Vinciguerra:2024}. The PDT-U is 
  the most complex model tested in Ref.~\cite{Vinciguerra:2024} and 
  is preferred by the Bayesian analysis.
\item PSR J0740+6620: We incorporate estimates from Salmi
  et~al.~\cite{Salmi:2024a}, Miller et~al.~\cite{Miller:2021}, and
  Riley et~al.~\cite{Riley:2021}. The representative estimates used 
  in this study were obtained from a joint NICER and XMM-Newton analysis 
  of the 2018-2022 dataset, based on the preferred ST-U model, which 
  provides a more comprehensive treatment of the background~\cite{Salmi:2024a}.
\item PSR J1231-1411: We consider three mass-radius estimates from
  the preferred PDT-U model with limited radius priors given in 
  Salmi et~al.~\cite{Salmi:2024b}. One estimate limited the radius 
  to be consistent with the previous observational constraints and 
  EoS analyses, and the other two estimates used an uninformative prior, 
  however, the radius was limited to ranges 10-14~km and 8-16~km, respectively. 
  It is important to note, however, that since the radius could not be 
  fully constrained independently in Ref.~\cite{Salmi:2024b}, the possibility 
  of better solutions deviating significantly from the above estimates 
  cannot be excluded.
\end{itemize}
By combining these observations, we establish three astrophysical
scenarios, as detailed in Table~\ref{table:data}. This approach
enables us to examine different possibilities within the framework of
our hybrid EoS models and the mass-twin hypothesis. Notably, the
ellipses derived for PSR J0437 significantly overlap with the
inferences from GW170817~\cite{LVScientific:2018}, further reinforcing
the selection of EoS based solely on gravitational wave data.

We do not consider the mass and radius estimates for the central compact object 
in HESS J1731-347, which was characterized as a very low-mass compact star with 
$M=0.77_{-0.17}^{+0.20} M_{\odot}$ and $R=10.4_{-0.78}^{+0.86}$~km in Doroshenko 
et al.~\cite{Doroshenko:2022}. From the theoretical point of view the options to 
explain this object have been extensively discussed recently and require some 
assumptions about properties of matter which are difficult to reconcile with the 
conventional models of dense matter, see for discussions 
Refs.~\cite{Lijj:2023b,Brodie:2023,Kubis:2023,Sagun:2023,Gao:2024,Lijj:2024,Veselsky:2024}. 
From the observational point of view, 
the analysis of Doroshenko et al.~\cite{Doroshenko:2022} has been questioned, 
in particular their use of the object's distance and uniform-temperature carbon 
atmosphere model, see  Ref.~\cite{Alford:2023}. It has been also argued that such 
a model does not perform well on longer-exposure XMM-Newton data from 2014 for the 
same source. Instead, a model featuring two hot regions emitting blackbody radiation
from the surface of a star of mass $\sim 1.4 M_{\odot}$, provides a better description 
of this spectrum~\cite{Alford:2023}. Consequently, there seems to be no consensus 
from the observational modeling point of view.

Additionally, we note that the companion of the ``black widow" pulsar PSR J0952-0607 
has been detected in the Milky Way. This pulsar, with a spin frequency of 707 Hz, 
has a mass estimate of $M=2.35_{-0.17}^{+0.17} M_{\odot}$ (68 \% credible interval)~\cite{Romani:2022}. 
Assuming uniform rotation, the increase in the maximum mass configuration due to 
rotation is approximately $0.05 M_{\odot}$~\cite{Lijj:2023d}. Thus, the lower limit 
on the mass of J0952 is comparable to that of J0740 at 95\% confidence, around 
$\sim$ $2.01 M_{\odot}$. This provides no additional constraints on our EoS models.

\subsection{Scenario A}
Figure~\ref{fig:MR_a} shows the \MR diagrams for hybrid EoS models
with four representative nucleonic EoS which are combined with the
quark matter EoS specified by two values of the speed of sound
$c_{\mathrm{s}}^2=1$ and $2/3$. Panel (a) is for stiff nucleonic EoS 
which demands a strong first-order phase transition with
$M_{\text {max}}^{\mathrm{H}} \lesssim 1.5\,M_{\odot}$ in order to be
consistent with the NICER inference for PSR J0437. Panel (b) uses
soft nucleonic EoS which could match PSR J0437's
inference without a phase transition to quark matter, but still, as an
alternative, may allow for phase transitions featuring twin
configurations.

\begin{table}[tb]
\centering
\setlength{\tabcolsep}{4.0pt}
\begin{tabular}{ccccccc}
\hline\hline
\multirow{2}*{EoS}& $\epsilon_{\rm{tr}}$ & $\Delta\epsilon/\epsilon_{\rm{tr}}$ &
$M^{\rm{H}}_{\rm{max}}$ & $M^{\rm{Q}}_{\rm{max}}$ & 
$\Delta M_{\rm{twin}}$ & $\Delta R_{\rm{twin}}$ 
\\
&[MeV/fm$^{3}$] & &$[M_{\odot}]$ &$[M_{\odot}]$ &
$[M_{\odot}]$ &[km] 
\\ 
\hline
DDLS100& 300.570 & 1.0615 & 1.30 & 2.10 & 0.0748 & 1.84 \\ 
       &         & 0.8881 & 1.30 & 2.20 & 0.0397 & 1.22 \\
       & 334.452 & 1.0055 & 1.50 & 2.05 & 0.0796 & 1.94 \\
       &         & 0.8349 & 1.50 & 2.15 & 0.0389 & 1.23 \\       
DDLS80 & 300.180 & 0.8764 & 1.20 & 2.20 & 0.0307 & 1.01 \\ 
       &         & 0.7249 & 1.20 & 2.30 & 0.0088 & 0.44 \\
       & 332.241 & 0.9100 & 1.40 & 2.10 & 0.0467 & 1.31 \\
       &         & 0.7506 & 1.40 & 2.20 & 0.0162 & 0.67 \\     
DDLS60 & 323.965 & 0.9362 & 1.30 & 2.10 & 0.0428 & 1.14 \\ 
       &         & 0.7739 & 1.30 & 2.20 & 0.0149 & 0.59 \\
       & 356.305 & 0.8252 & 1.50 & 2.10 & 0.0323 & 0.99 \\
       &         & 0.6735 & 1.50 & 2.20 & 0.0049 & 0.34 \\   
DDLS40 & 310.074 & 0.8291 & 1.20 & 2.20 & 0.0187 & 0.61 \\ 
       &         & 0.6817 & 1.20 & 2.30 & 0.0023 & 0.15 \\
       & 338.751 & 0.8856 & 1.40 & 2.10 & 0.0358 & 0.98 \\
       &         & 0.7283 & 1.40 & 2.20 & 0.0089 & 0.41 \\        
\hline\hline
\end{tabular}
\caption{
Parameters of the hybrid EoS models used in this work, and the 
characteristics of their mass-radius curves. All have the maximum 
sound speed $\csq = 1$ in the quark phase. The last two columns show 
the ranges of mass and radius within which twin configurations exist.}
\label{table:Parameters}
\end{table}

\begin{table}[tb]
\centering
\setlength{\tabcolsep}{4.0pt}
\begin{tabular}{ccccccc}
\hline\hline
\multirow{2}*{EoS}& $\epsilon_{\rm{tr}}$ & $\Delta\epsilon/\epsilon_{\rm{tr}}$ &
$M^{\rm{H}}_{\rm{max}}$ & $M^{\rm{Q}}_{\rm{max}}$ & 
$\Delta M_{\rm{twin}}$ & $\Delta R_{\rm{twin}}$ 
\\
&[MeV/fm$^{3}$] & &$[M_{\odot}]$ &$[M_{\odot}]$ &
$[M_{\odot}]$ &[km] 
\\ 
\hline
DDLS100& 300.570 & 0.7823 & 1.30 & 2.00 & 0.0225 & 0.92 \\ 
       &         & 0.6298 & 1.30 & 2.10 & 0.0015 & 0.21 \\
       & 334.452 & 0.6898 & 1.50 & 2.00 & 0.0124 & 0.69 \\
       &         & 0.5467 & 1.50 & 2.10 & -      & -    \\
                 
DDLS80 & 300.180 & 0.7641 & 1.20 & 2.00 & 0.0148 & 0.64 \\
       &         & 0.6125 & 1.20 & 2.10 & -      & -    \\ 
       & 332.241 & 0.6735 & 1.40 & 2.00 & 0.0059 & 0.41 \\
       &         & 0.5313 & 1.40 & 2.10 & -      & -    \\ 
                  
DDLS60 & 323.965 & 0.6851 & 1.30 & 2.00 & 0.0045 & 0.29 \\ 
       &         & 0.5414 & 1.30 & 2.10 & -      & -    \\  
       & 356.305 & 0.6198 & 1.50 & 2.00 & 0.0006 & 0.14 \\
       &         & 0.4826 & 1.50 & 2.10 & -      & -    \\ 
              
DDLS40 & 310.074 & 0.7256 & 1.20 & 2.00 & 0.0063 & 0.31 \\ 
       &         & 0.5777 & 1.20 & 2.10 & -      & -    \\   
       & 338.751 & 0.6573 & 1.40 & 2.00 & 0.0014 & 0.18 \\
       &         & 0.5166 & 1.40 & 2.10 & -      & -    \\       
\hline\hline
\end{tabular}
\centering
\caption{
Same as Table~\ref{table:Parameters} but for models have the intermediate 
sound speed $\csq = 2/3$ in the quark phase.}
\label{table:Parameters_b}
\end{table}

Figure~\ref{fig:MR_a} demonstrates that Scenario A contains
models that are consistent with all four of the \MR measurements. 
Each of the curves has a hadronic branch that reaches
within the 95\% contours for PSR J0030, and a hybrid branch that
passes through the 95\% contours for J0437 and J0740. 
(Note that the three contours for PSR J1231 are distinctly different,
the stiff hadronic models meet those two larger-radius contours, 
while soft hadronic models meet those two smaller-radius contours instead.)
In several cases the \MR curves are even consistent with the 
68\% contours.

For the models with higher transition density, corresponding to
$M_{\max }^{\mathrm{H}}=1.5\,M_{\odot}$ or $1.4\,M_{\odot}$ it is
possible that J0030 and J0437 could be hadronic-hybrid twins.
For models with a lower transition density
($M_{\max }^{\mathrm{H}} \lesssim 1.3\,M_{\odot}$) the mass range
where twins exist is below the mass interval of PSR J0437, so that
PSR J0030 could be a hadronic or hybrid star, while J0437 must be a
hybrid star in this case.

Tables~\ref{table:Parameters} and~\ref{table:Parameters_b} present the
parameters of the used hybrid EoS models with $\csq=1$ and $2/3$,
respectively, in quark phase and characteristics of the corresponding
\MR diagrams, i.e., the values of
$M_{\max }^{\mathrm{H}}, M_{\max }^{\mathrm{Q}}$, and ranges of mass
and radius that characterize twin configurations,
$\Delta M_{\text {twin}}=M_{\text {max}}^{\mathrm{H}}-M_{\text
  {min}}^{\mathrm{Q}}$, and $\Delta R_{\mathrm{twin}}$ the radius
difference between the $M_{\max }^{\mathrm{H}}$ hadronic star and that
of the hybrid counterpart with the same mass. Note that the case
$\csq=1$ allows us to establish the maximum range where twin
configurations exist in our setup.

\subsection{Scenario B} 
\begin{figure*}[b]
\centering
\includegraphics[width = 1.0\textwidth]{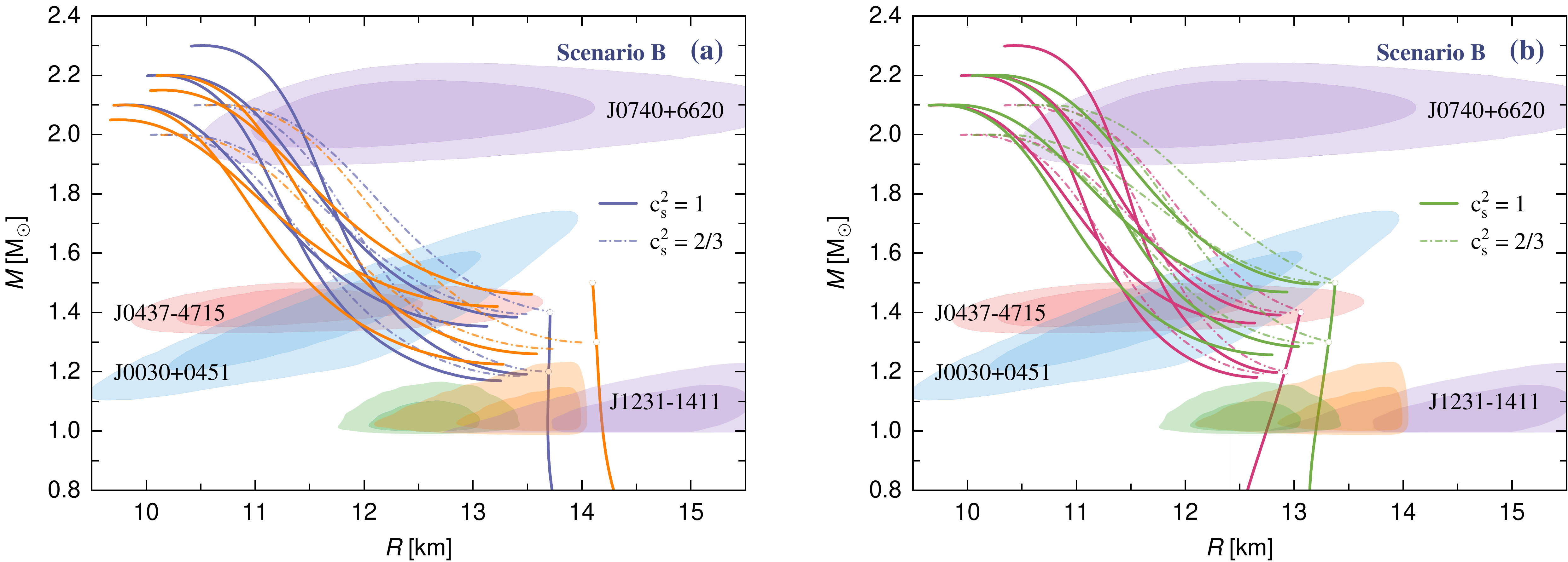}
\caption{Same as Figure~\ref{fig:MR_a}, but for scenario B, where 
the light blue ellipse corresponds to ST+PDT analysis 
for PSR J0030~\cite{Vinciguerra:2024}.}
\label{fig:MR_b}
\end{figure*}
This is the scenario where the \MR ellipses for canonical mass stars
PSR J0030 and J0437 are maximally overlapping; see~Figure~\ref{fig:MR_b}. 
In this case, hadronic stars are consistent with data only at $2\sigma$ 
($95\%$ confidence) level, and only for soft hadronic matter (panel b). 
If the hadronic EoS is stiff then both stars must be hybrid which means 
they may have hadronic twins with significantly larger radii $R\ge 13.7$~km
(the $2\sigma$ upper limit for J0437) in our example, see panel (a).

\subsection{Scenario C} 
\begin{figure*}[tb]
\centering
\includegraphics[width = 1.0\textwidth]{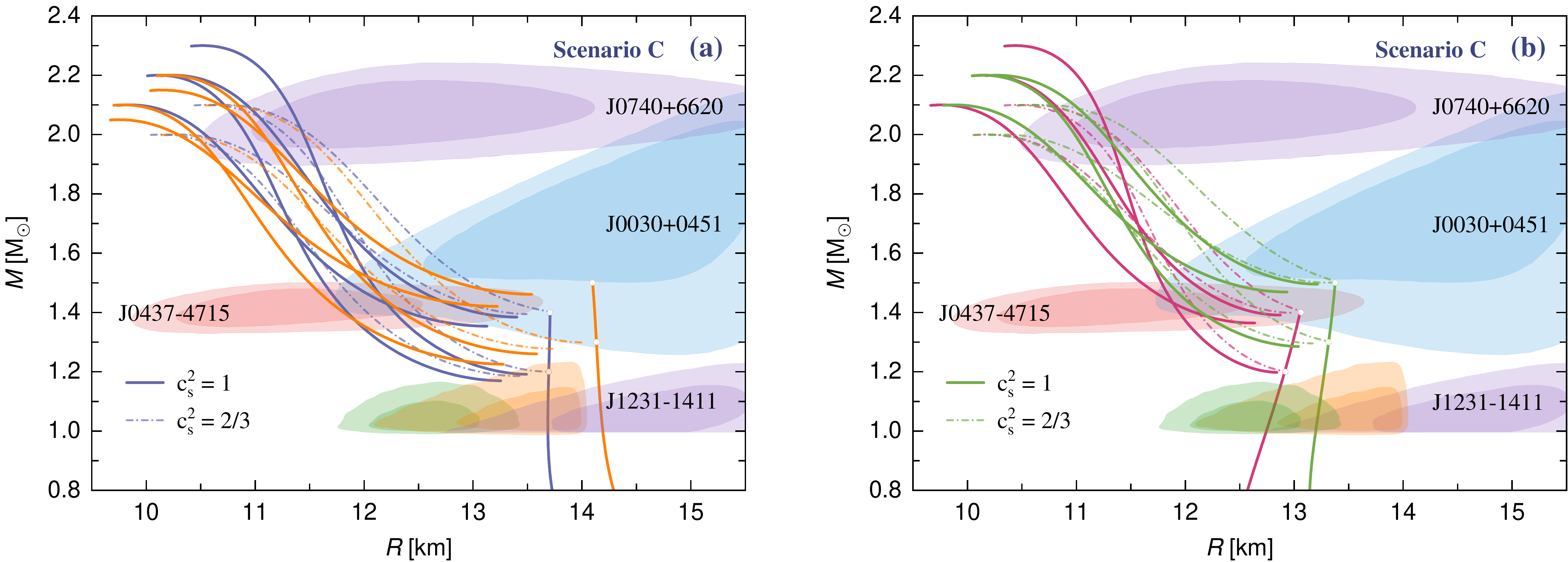}
\caption{Same as Figure~\ref{fig:MR_a}, but for scenario C, where 
the light blue ellipse corresponds to PDT-U analysis 
for PSR J0030~\cite{Vinciguerra:2024}.}
\label{fig:MR_c}
\end{figure*}
This is the scenario where the \MR ellipses for PSR J0030 and J0437
overlap only at $2\sigma$ ($95\%$ confidence), see~Figure~\ref{fig:MR_c}. 
This scenario can be viewed as a more extreme version of Scenario A, 
therefore, the conclusions drawn above apply to this scenario too. 
Nevertheless, there are some quantitative differences. PSR J0030 has 
now a large radius $R>12$~km, which excludes models of EoS with soft 
hadronic matter which produce low values of $M_{\max }^{\mathrm{H}}$ or 
$M_{\max }^{\mathrm{Q}}$ with $R_{1.4}<12$~km (the $2\sigma$ lower limit 
for J0030), where $R_{1.4}$ is the radius of $1.4\,M_{\odot}$ star. 
This is the case, for example, for the EoS based on DDLS-60 
($M_{\max }^{\mathrm{H}} =1.3\,M_{\odot}$, 
$M_{\max }^{\mathrm{Q}} =2.1\,M_{\odot}$ and $\csq = 1$) 
and DDLS-40 ($M_{\max }^{\mathrm{H}} =1.2\,M_{\odot}$,
$M_{\max }^{\mathrm{Q}} =2.0\,M_{\odot}$ and $\csq = 2/3$).

\begin{figure*}[b]
\centering
\includegraphics[width = 1.0\textwidth]{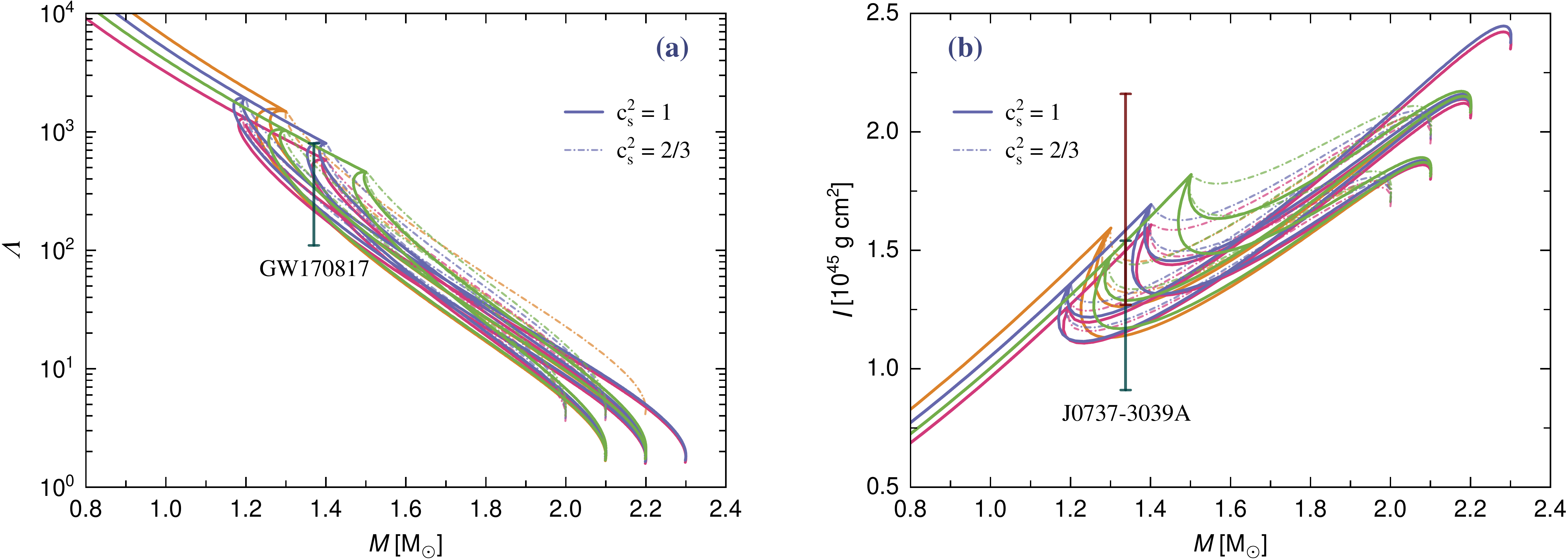}
\caption{ Mass-tidal deformability (left panel) and mass-moment of 
inertia (right panel) relations for EoS models. The 90\% confidence 
ranges for a $1.362\,M_{\odot}$ star deduced from the analysis of 
GW170817~\cite{LVScientific:2018} (left panel) and for $1.338\,M_{\odot}$ 
PSR J0737-3039 A from radio observations~\cite{Kramer:2021PRX} (right panel) 
are shown with the vertical error bars.}
\label{fig:MLambda}
\end{figure*}

\subsection{Tidal deformability and moment of inertia}
In this section, we consider two additional integral parameters of
hybrid stars that have observational significance - the tidal
deformability and moment of inertia.
Figure~\ref{fig:MLambda} (a) shows the dimensionless tidal
deformability vs mass relation for our models. It is seen that the
models discussed in this work satisfy the constraint placed for a
$1.362\,M_{\odot}$ star from the analysis of the GW170817
event~\cite{LVScientific:2018}. It is also seen that the softer EoS 
safely passes through the required range for $\Lambda_{1.362}$. 
Furthermore, for hybrid models for which 
$M_\text{min}^{\rm Q}\le 1.4\,M_{\odot}$
the transition to quark matter improves the agreement with the data.

Figure~\ref{fig:MLambda} (b) shows the moment of inertia predicted by
EoS models for $1.338\,M_{\odot}$ star. A motivation for considering 
the moment of inertia comes from the observation of binary
PSR J0737-3039~A~\cite{Burgay:2003Nature,Kramer:2021PRX}, which shows
changes in orbital parameters, such as the orbit inclination (the
angle between the orbital plane and observer's line of sight) and the
preriastron position~\cite{Lattimer:2005ApJ}. It is evident from the
figure that our models of hybrid stars are broadly consistent with the
constraint $0.91 \le I_{1.338} \le 2.16$ inferred from 16-yr data span
reported in Ref.~\cite{Kramer:2021PRX}, where $I_{1.338}$ is the
moment of inertia of a neutron star with $1.338\,M_{\odot}$ mass in
units of $10^{45} {\rm g\,cm}^2$.

\section{Conclusions}
\label{sec:Conclusions}
Motivated by the recent (re)analysis of the data on two X-ray emitting
pulsars PSR J0030+0451 and J0470+6620 as well as new results on PSR
J0437-4715 and J1231-1411 we compared the new ellipses in the \MR diagram 
for these pulsars with our models of hybrid stars, which are based on 
CDF EoS for nucleonic matter at low densities and quark matter EoS,
parametrized by speed of sound, at higher densities. These models are
also validated by comparisons of their predicted tidal deformabilities
with the observations of GW170817 and predicted moment of inertia with
the constraints for PSR J0737-3039 A, see~Figure~\ref{fig:MLambda}.

In more detail, we considered three possible scenarios A, B and C
which correspond to the three mass and radius estimates taken from a
reanalysis of 2017-2018 data~\cite{Vinciguerra:2024} for PSR
J0030+0451.  These include an improved NICER-only ST+PST analysis and
two joint NICER-XMM-Newton models (ST+PDT and PDT-U), with the
Bayesian-preferred PDT-U being the most complex. We then examined the
consistency of the three scenarios with our models with a special
focus on the possibility of the formation of twin stars.  We find that
in two scenarios (A and C), where the ellipses for the canonical mass
($\sim 1.4\,M_{\odot}$) stars J0030+0451 and J0437-4715 exhibit the
maximal mismatch, the potential difference in the radii of these
stars within these scenarios is naturally explained by the presence of
twin stars.

To conclude, the ability of the hybrid star models to explain
observational data from multiple sources (X-ray pulsars and
gravitational waves) encourages further refinement of theoretical
models, potentially leading to more accurate descriptions of neutron
star interiors. This can be only achieved through combining
theoretical models (for example using the classes of models discussed
here) with observational data, which is expected to improve over time.

\section*{Acknowledgments}  
J.~L. is supported by the National Natural Science 
Foundation of China under Grant No. 12105232 and 
the Fundamental Research Funds for the Central
Universities under Grant No. SWU-020021.
A.~S. is funded by Deutsche Forschungsgemeinschaft 
Grant No. SE 1836/5-3 and the Polish NCN Grant 
No. 2023/51/B/ST9/02798. 
M.~A. is partly supported  by the U.S. Department of 
Energy, Office of Science, Office of Nuclear Physics 
under Award No. DE-FG02-05ER41375.


\providecommand{\href}[2]{#2}\begingroup\raggedright
\endgroup

\end{document}